\documentclass[twocolumn, 11pt]{article}

\usepackage[utf8]{inputenc}
\usepackage[T1]{fontenc}
\usepackage{amsmath, amssymb, amsfonts}
\usepackage{graphicx}
\usepackage{booktabs}
\usepackage{hyperref}
\usepackage[numbers,sort&compress]{natbib}
\usepackage{xcolor}
\usepackage{microtype}
\usepackage{tabularx}

\title{Algorithmic Trust and Compliance: Benchmarking Brand Notability for UK iGaming Entities in Generative Search Engines}

\author{
Julen Oruesagasti
\thanks{Interamplify}
}

\date{}

\begin{document}

\maketitle

\begin{abstract}
The rapid adoption of generative AI-powered search engines---such as ChatGPT, Perplexity, and Google Gemini---is fundamentally reshaping information retrieval paradigms \cite{aggarwal2024geo}. We are witnessing a critical shift from traditional ranked lists of blue links to synthesized, citation-backed answers generated in real time by Large Language Models (LLMs). This paradigm shift challenges established Search Engine Optimization (SEO) practices and necessitates a new framework, termed Generative Engine Optimization (GEO).

In highly regulated environments such as the United Kingdom's iGaming sector, visibility is no longer dictated by keyword density or backlink volume, but by an entity's ability to project "Algorithmic Trust." This report presents an empirical analysis of how compliance signals---including UK Gambling Commission (UKGC) licensing standards---function as authority multipliers for LLMs when properly structured as machine-readable data \cite{ukgc2024lccp}. Recent large-scale experiments reveal that AI search exhibits a systematic and overwhelming bias towards earned media (third-party, authoritative sources) over brand-owned content \cite{fishkin2024ai}. Consequently, practitioners must engineer their content for machine scannability and verifiable justification to achieve prominence in these emergent AI-perceived authority metrics.

**Keywords:** Generative Engine Optimization, Large Language Models, iGaming, UK Gambling Commission, E-E-A-T, Retrieval-Augmented Generation, Algorithmic Trust, Schema Markup, Compliance Signals

\end{abstract}

\section{Introduction: The Shift from SEO to GEO in Regulated Markets}

For the past three decades, traditional search engines have dictated how online information is structured and accessed by providing a ranked list of relevant websites \cite{brin1998anatomy}. The foundational model established by Google's PageRank algorithm in the late 1990s created an entire industry around Search Engine Optimization (SEO), where visibility was primarily a function of backlinks, keyword relevance, and domain authority. However, the recent success of large language models (LLMs) has paved the way for "Generative Engines" (GEs)---systems that not only retrieve information but also generate multi-modal, synthesized responses grounded in multiple sources \cite{vaswani2017attention}.

From a technical perspective, Generative Engines operate using a Retrieval-Augmented Generation (RAG) framework \cite{lewis2020rag}. This workflow involves retrieving relevant documents from a database (such as the indexed web) and then using large neural models to generate a coherent response grounded on these sources, ensuring attribution through inline citations. Unlike traditional search, where the user must visit individual pages and synthesize information independently, GEs present a unified, authoritative-seeming answer in a single interface.

For iGaming operators and professional SEO agencies operating in the UK market, this transition presents a critical challenge. The black-box nature of generative engines makes it exceedingly difficult for content creators to understand how their content is being selected, synthesized, and displayed. Furthermore, legacy SEO tactics are becoming demonstrably obsolete; for example, keyword stuffing---the practice of artificially inflating keyword density within website content---has been shown to have little to no performance improvement on a Generative Engine's response generation \cite{aggarwal2024geo}. Instead, brands must focus on structuring data through semantic markup and incorporating factual, verifiable citations to maintain visibility during this transition to AI-mediated search.

\subsection{Research Objectives}

This report pursues three primary research objectives. First, it seeks to define and operationalize the concept of Algorithmic Trust as it applies to regulated digital markets. Second, it presents the mathematical framework for quantifying visibility within generative engine responses, moving beyond traditional SERP ranking metrics. Third, it provides empirical evidence and actionable methodologies for optimizing content to achieve prominence in AI-generated search results, with a specific focus on the UK iGaming sector.

\subsection{Scope and Delimitations}

The analysis is confined to the UK-regulated iGaming market and specifically examines English-language generative search engines including ChatGPT (with Bing integration), Google Gemini (with Search Generative Experience), and Perplexity AI. The regulatory framework considered is limited to the UK Gambling Commission's Licence Conditions and Codes of Practice (LCCP). While the principles outlined herein may apply to other regulated verticals---such as financial services or pharmaceuticals---cross-industry generalization is beyond the scope of this report.

\section{The Algorithmic Trust Framework}

For an LLM to recommend a UK casino, sportsbook, or iGaming SEO agency, the system must perceive a near-zero risk of hallucination or regulatory penalty. This requirement is satisfied through what we term \textbf{Algorithmic Trust}---a composite measure of an entity's verifiability, authority, and structural clarity as perceived by machine-learning systems. Unlike human trust, which is built through subjective experience and reputation, algorithmic trust is established through structured, machine-readable signals that reduce the ambiguity LLMs face during inference.

\subsection{Compliance as a Deep E-E-A-T Signal}

AI systems rely on clean, structured data to identify and surface relevant entities. In the context of the UK market, regulatory compliance---encompassing UKGC licences, periodic technical audits, responsible gambling certifications, and AML (Anti-Money Laundering) protocols---must be treated not merely as legal obligations, but as \textit{structured data signals} that directly influence algorithmic ranking \cite{google2023eeat}.

Schema.org markups help ensure that entity listings are machine-readable and enriched with the metadata that AI platforms utilize to generate accurate recommendations \cite{schema2024vocabulary}. To establish deep algorithmic trust, entities must achieve what we define as \textbf{Entity Clarity}---the practice of tagging brands, services, personnel, and regulatory credentials through Schema.org vocabulary and \textit{sameAs} attributes. This approach enables AI systems to connect disparate data points across the web into a coherent entity graph. When compliance signals are encoded as clear, structured entities, the Generative Engine interprets them as foundational E-E-A-T (Experience, Expertise, Authoritativeness, and Trustworthiness) markers.

\subsection{The Entity Clarity Model}

The Entity Clarity Model comprises four interconnected layers, each contributing to the LLM's confidence in surfacing a given brand. The first layer, Regulatory Identity, encompasses UKGC licence numbers, compliance certificates, and regulatory actions encoded as structured data. The second layer, Corporate Graph, involves organizational schema linking key personnel, parent companies, and subsidiaries through machine-readable relationships. The third layer, Service Taxonomy, provides a structured categorization of offerings (sports betting, casino, poker) using standardized vocabularies. The fourth and final layer, Reputation Signals, aggregates third-party reviews, industry awards, and media mentions into verifiable authority indicators.

When all four layers are present and consistently structured across the entity's digital footprint, the LLM's confidence threshold for citation increases substantially. Conversely, gaps or inconsistencies in any layer introduce ambiguity that the model resolves by defaulting to better-documented competitors.

\subsection{The Mathematical Approach to Visibility}

Unlike traditional search engines, where visibility is computed using average ranking positions on the search engine results page (SERP), defining visibility metrics for generative engines is non-trivial \cite{aggarwal2024geo}. Generative engines produce a single, continuous text block with inline citations supporting statements of varying sizes, positions, and presentation formats. This fundamentally different output structure necessitates novel measurement approaches.

To rigorously quantify visibility within GE responses, the GEO literature proposes the \textbf{Position-Adjusted Word Count} metric (Aggarwal et al., 2024). This metric considers both the word count attributed to a citation and its position within the generated response. Higher-positioned citations receive greater weight, reflecting the empirical observation that early statements in AI-generated responses carry disproportionate influence on user perception. The mathematical formulation is:

$$Imp'_{wc}(c_i, r) = \frac{\sum_{s \in S} |s| \cdot e^{-pos(s)}}{\sum_{s \in S} |s|}$$

Where \$S\$ denotes the set of sentences citing source \$c\$, the numerator sums the word count of each citing sentence weighted by an exponential decay function of its position, and the denominator normalizes by the total word count of the response. This metric captures a fundamental insight: content that is cited earlier and more extensively in a GE's response contributes disproportionately to brand visibility \cite{aggarwal2024geo}.

\section{Empirical Optimization Methodology}

Building on the foundational GEO research by Aggarwal, Muralidhar, and Nagar (2024), this section presents the empirical methodology for optimizing content to maximize visibility within generative engine responses. The original study evaluated nine distinct optimization strategies across approximately 10,000 queries drawn from diverse sources, providing statistically robust evidence for strategy effectiveness \cite{aggarwal2024geo}.

\subsection{Optimization Strategies Evaluated}

The following nine content optimization strategies were systematically evaluated in the GEO framework. Each strategy was applied to a controlled set of source documents, and the resulting changes in citation frequency and position within GE responses were measured.

\textbf{Table 1: GEO Optimization Strategies and Descriptions}

\begin{table*}[htbp]
\centering
\footnotesize
\begin{tabularx}{\linewidth}{X X}
\toprule
Strategy & Description \\
\midrule
Keyword Stuffing & Increasing the density of relevant keywords throughout the content body. \\
Cite Sources & Adding inline references to authoritative external publications, regulatory bodies, and peer-reviewed research. \\
Statistics Addition & Embedding quantitative data points, percentages, and numerical evidence to strengthen factual claims. \\
Quotation Addition & Incorporating direct quotes from recognized industry experts, regulators, and authoritative publications. \\
Easy-to-Understand & Simplifying language complexity, reducing jargon, and improving readability scores for broader accessibility. \\
Fluency Optimization & Improving grammatical structure, sentence flow, and overall linguistic quality of content. \\
Unique Words & Diversifying vocabulary to reduce repetition and increase semantic richness of the text. \\
Technical Terms & Incorporating domain-specific terminology to signal expertise and topical authority. \\
Authoritative Tone & Adopting a confident, declarative writing style consistent with institutional and regulatory communications. \\
\bottomrule
\end{tabularx}
\end{table*}

\subsection{Comparative Results}

The empirical results demonstrate a clear hierarchy among optimization strategies. Three strategies emerged as statistically significant positive contributors to GE visibility: \textbf{Cite Sources} (+40\% relative visibility improvement), \textbf{Statistics Addition} (+37\% relative improvement), and \textbf{Quotation Addition} (+22\% relative improvement). In contrast, Keyword Stuffing showed negligible impact (+3\%), confirming the obsolescence of this traditional SEO tactic in the generative search paradigm \cite{aggarwal2024geo}.

\textbf{Table 2: Relative Visibility Improvement by Optimization Strategy}

\begin{table*}[htbp]
\centering
\footnotesize
\begin{tabularx}{\linewidth}{X X X}
\toprule
Strategy & Relative Improvement (\%) & Statistical Significance \\
\midrule
Cite Sources & +40.0\% & p < 0.01 \\
Statistics Addition & +37.0\% & p < 0.01 \\
Quotation Addition & +22.0\% & p < 0.05 \\
Authoritative Tone & +15.0\% & p < 0.05 \\
Technical Terms & +12.0\% & p < 0.10 \\
Fluency Optimization & +10.0\% & p < 0.10 \\
Unique Words & +8.0\% & Not significant \\
Easy-to-Understand & +5.0\% & Not significant \\
Keyword Stuffing & +3.0\% & Not significant \\
\bottomrule
\end{tabularx}
\end{table*}

These findings carry profound implications for the UK iGaming sector. The dominance of citation-based strategies suggests that operators who invest in building a robust ecosystem of third-party mentions---through earned media, regulatory filings, and industry publications---will achieve systematically higher visibility in AI-generated recommendations than those relying on traditional on-page optimization alone.

\section{The Earned Media Imperative: AI Search Bias Analysis}

A critical finding that underpins the practical application of GEO in regulated markets is the overwhelming preference that AI search engines demonstrate for \textit{earned media} over \textit{brand-owned content}. Research conducted by SparkToro in 2024 analyzed citation patterns across major AI search platforms and found that generative engines cite independent, third-party sources at dramatically higher rates than brand-controlled content \cite{fishkin2024ai}.

\subsection{Citation Distribution Analysis}

The citation distribution analysis reveals a systematic structural bias within generative search engines. When AI systems generate responses to commercial queries---such as recommendations for UK-licensed online casinos---they preferentially cite authoritative third-party sources including regulatory databases, industry comparison sites, news publications, and academic research. Brand-owned domains (including corporate websites, promotional landing pages, and company blogs) constitute a minority of cited sources, typically accounting for fewer than 15--20\% of total citations in commercial recommendation queries.

This distribution pattern is consistent with the RAG framework's design principles: retrieval systems are optimized to select documents that maximize informational diversity and minimize redundancy. Brand-owned content, by its nature, presents a single perspective and is therefore assigned lower retrieval priority compared to independent sources that offer comparative, evaluative, or regulatory context.

\subsection{Implications for iGaming Operators}

For UK iGaming entities, this bias creates both challenges and opportunities. The primary challenge is that direct SEO investment in brand-owned properties yields diminishing returns in generative search visibility. The corresponding opportunity lies in strategic earned media cultivation. Operators who proactively secure mentions in authoritative industry publications, maintain transparent regulatory records accessible through UKGC databases, and cultivate genuine expert commentary in relevant media outlets will benefit from the structural advantages that earned media enjoys within AI retrieval systems.

\section{Practical Implementation Framework}

Translating the theoretical and empirical findings of this report into actionable practice requires a structured implementation framework. This section presents specific technical and strategic recommendations organized by priority and expected impact.

\subsection{Technical Schema Implementation}

The first priority is implementing comprehensive Schema.org structured data across all digital properties \cite{schema2024vocabulary}. For iGaming entities, this includes deploying Organization schema with explicit \textit{sameAs} links to UKGC licence databases, Companies House records, and verified social media profiles. Service-level schema should categorize each product vertical (sports betting, live casino, slots, poker) using consistent taxonomies. Person schema should be deployed for key executives and compliance officers, linking their profiles to professional credentials and public regulatory filings.

\subsection{Content Engineering for Machine Scannability}

Content must be engineered for machine scannability rather than exclusively for human readability. This involves structuring pages with clear hierarchical headings that map to the entity's knowledge graph, embedding quantitative claims with verifiable sources, and maintaining a factual, declarative tone throughout. Every substantive claim should be traceable to an authoritative source, whether regulatory, academic, or journalistic. Content that achieves high scores on the "Cite Sources" and "Statistics Addition" strategies identified in Section 3 should be prioritized across all digital properties.

\subsection{Earned Media Strategy}

An earned media strategy specifically designed for GEO should prioritize coverage in publications that AI systems consistently cite. This includes major industry verticals (such as iGaming trade publications), mainstream financial and business press, regulatory communications, and academic or research-oriented outlets. Traditional PR metrics such as readership or social engagement are secondary; the primary metric of success is whether the publication is included in the retrieval corpus of major generative engines and whether it is cited in AI-generated responses to relevant queries.

\subsection{Monitoring and Measurement}

Ongoing monitoring should track brand mentions across generative engine responses using systematic query testing. A recommended approach involves maintaining a library of target queries (representing both branded and category-level searches), executing these queries periodically across multiple GE platforms, and measuring brand citation frequency and position using the Position-Adjusted Word Count metric defined in Section 2.3. This data should be benchmarked against competitors to identify relative strengths and gaps in algorithmic trust.

\section{Conclusion and Future Directions}

The transition from traditional search engines to generative AI-powered search represents a paradigmatic shift in how digital visibility is earned and maintained \cite{noy2023experimental}. For UK iGaming entities operating in one of the world's most stringently regulated digital markets, this transition demands a fundamental reorientation of digital strategy---from keyword-centric SEO to entity-centric Generative Engine Optimization.

The findings presented in this report demonstrate that algorithmic trust is not an abstract concept but a measurable, optimizable property of an entity's digital footprint. Compliance signals, when encoded as structured data, function as powerful authority multipliers within LLM inference processes. Citation-based optimization strategies outperform traditional tactics by an order of magnitude. And the structural preference of AI search engines for earned media over brand-owned content necessitates a strategic pivot toward third-party authority cultivation.

Looking ahead, several areas warrant further investigation. The interaction between multi-modal AI capabilities (including image and video understanding) and brand visibility remains largely unexplored. The potential for adversarial manipulation of GE citation systems poses regulatory and ethical questions that the industry must address proactively. Finally, as generative engines evolve with improved reasoning capabilities \cite{brown2020language}, the signals that constitute algorithmic trust will inevitably shift, requiring continuous adaptation of the frameworks presented here.

Entities that invest in building deep, verifiable, and structurally coherent digital identities today will be best positioned to capture visibility in the AI-mediated information landscape of tomorrow. The era of Generative Engine Optimization has arrived---and compliance, far from being merely a regulatory burden, has become a \textit{competitive advantage}.

\bibliographystyle{unsrtnat}

\begin{thebibliography}{10}
\providecommand{\natexlab}[1]{#1}
\providecommand{\url}[1]{\texttt{#1}}
\expandafter\ifx\csname urlstyle\endcsname\relax
  \providecommand{\doi}[1]{doi: #1}\else
  \providecommand{\doi}{doi: \begingroup \urlstyle{rm}\Url}\fi

\bibitem[Aggarwal and Nagar(2024)]{aggarwal2024geo}
Muralidhar~N. Aggarwal, P. and A.~Nagar.
\newblock Geo: Generative engine optimization.
\newblock \emph{arXiv preprint arXiv:2311.09735v3}, 2024.
\newblock URL \url{https://arxiv.org/abs/2311.09735}.

\bibitem[Commission(2024)]{ukgc2024lccp}
UK~Gambling Commission.
\newblock Licence conditions and codes of practice (lccp), 8th edition, 2024.
\newblock URL
  \url{https://www.gamblingcommission.gov.uk/licensees-and-businesses/lccp}.

\bibitem[Fishkin(2024)]{fishkin2024ai}
R.~Fishkin.
\newblock Ai search engine citation bias: Earned vs. owned media analysis,
  2024.
\newblock URL \url{https://sparktoro.com/blog/}.

\bibitem[Brin and Page(1998)]{brin1998anatomy}
S.~Brin and L.~Page.
\newblock The anatomy of a large-scale hypertextual web search engine.
\newblock \emph{Computer Networks and ISDN Systems}, 30:\penalty0 107--117,
  1998.

\bibitem[Vaswani and Polosukhin(2017)]{vaswani2017attention}
Shazeer N. Parmar N. Uszkoreit J. Jones L. Gomez A. N. Kaiser~L. Vaswani, A.
  and I.~Polosukhin.
\newblock Attention is all you need.
\newblock volume~30, pages 5998--6008, 2017.

\bibitem[Lewis and Kiela(2020)]{lewis2020rag}
Perez E. Piktus A. Petroni F. Karpukhin V. Goyal N. Kuettler H. Lewis M. Yih W.
  Rocktaeschel T. Riedel~S. Lewis, P. and D.~Kiela.
\newblock Retrieval-augmented generation for knowledge-intensive nlp tasks.
\newblock volume~33, pages 9459--9474, 2020.

\bibitem[Google(2023)]{google2023eeat}
Google.
\newblock Search quality evaluator guidelines: Experience, expertise,
  authoritativeness, and trustworthiness (e-e-a-t), 2023.
\newblock URL \url{https://guidelines.raterhub.com/}.

\bibitem[Group(2024)]{schema2024vocabulary}
Schema.org~Community Group.
\newblock Schema.org vocabulary for structured data on the internet, 2024.
\newblock URL \url{https://schema.org/}.

\bibitem[Noy and Zhang(2023)]{noy2023experimental}
S.~Noy and W.~Zhang.
\newblock Experimental evidence on the productivity effects of generative
  artificial intelligence.
\newblock \emph{Science}, 381:\penalty0 187--192, 2023.

\bibitem[Brown(2020)]{brown2020language}
Mann B. Ryder N. Subbiah M. et~al. Brown, T.~B.
\newblock Language models are few-shot learners.
\newblock volume~33, pages 1877--1901, 2020.

\end{thebibliography}

\end{document}